\documentstyle[preprint,prl,aps,graphicx]{revtex}

\def\beq{\begin{equation}}
\def\eeq{\end{equation}}
\def\beqa{\begin{eqnarray}}
\def\eeqa{\end{eqnarray}}

\def\za{\alpha}
\def\zb{\beta}
\def\lsim{\mathrel{\raise.3ex\hbox{$<$\kern-.75em\lower1ex\hbox{$\sim$}}} }
\def\gsim{\mathrel{\raise.3ex\hbox{$>$\kern-.75em\lower1ex\hbox{$\sim$}}} }

\begin{document}
\draft \preprint{{\vbox{\hbox{IPAS-HEP-k013}\hbox{NSC-NCTS-010316}
\hbox{Mar 2001}
\hbox{rev. Jul 2001}}}}

\title{Muon Anomalous Magnetic Moment, Two-Higgs-Doublet Model, and
Supersymmetry }
\author{\bf Kingman Cheung$^1$
$\!$\footnote{E-mail: cheung@phys.cts.nthu.edu.tw},
Chung-Hsien Chou$^2$ $\!$\footnote{E-mail: chouch@phys.sinica.edu.tw},
and Otto C. W. Kong$^2$
$\!$\footnote{Permanent address since Aug. 2001 : Department of Physics, 
	National Central University, Chung-li, TAIWAN 32054. E-mail: otto@phy.ncu.edu.tw}
}
\address{$^1$National Center for Theoretical Science, National Tsing Hua
University, Hsinchu, Taiwan \\
$^2$Institute of Physics, Academia Sinica,
Nankang, Taipei, Taiwan 11529}
\maketitle

\begin{abstract}
The recent measurement of the muon anomalous magnetic moment $a_\mu$ shows a
$2.6\sigma$ deviation from the standard model value. We show that it puts
strong constraints on the parameter space of the two-Higgs doublet model
(2HDM) II. The dominant contribution of the Higgs bosons comes at the two-loop level,
and in order to explain the data it favors a pseudoscalar $A$ with a
light mass range and a large $\tan\!\beta$. At 95\% C.L., the upper limit for
$m_{\!\scriptscriptstyle A}$ is $29\,(55)\,(85)$ GeV for $\tan\!\beta=30\,(45)
\,(60)$, and $\tan\!\beta$ is bounded below at 17. This is in sharp contrast
to the conclusion one draws from considering one-loop Higgs contributions alone.
Finally, we also discuss the role of the Higgs contributions in the
minimal supersymmetric standard model.
\end{abstract}

\newpage

The recent result on the measurement of muon anomalous magnetic
moment ($a_\mu$) by the experiment E821\cite{E821} at Brookhaven National
Laboratory has reduced the error to a very small level.
Comparing the data with the prediction from the Standard Model
(SM)\cite{CM}, one gets the deviation
\begin{equation} \label{exp}
\Delta a_\mu \equiv a_\mu^{\rm exp} -a_\mu^{\rm SM} =
426\, (165) \; \times \; 10^{-11} \;,
\end{equation}
which may suggest the presence of contributions from physics beyond the SM.
Taking the above numbers at face value, the range of $\Delta a_\mu$ at 95\% C.L.
($\pm 1.96\sigma$) is given by
\begin{equation}
\label{95}
10.3\times 10^{-10} < \Delta a_\mu < 74.9 \times 10^{-10} \;.
\end{equation}

Most of the extensions of the SM start with an extended Higgs sector, the
simplest of which is the two-Higgs-doublet model (2HDM).  In particular,
the more interesting model II\cite{guide} shares the same Higgs structure
as the minimal supersymmetric standard model (MSSM).
Hence, going beyond the SM to look for extra contributions to $a_\mu$,
the 2HDM should be among the first to be examined seriously.

In this Rapid Communication, we investigate the contributions from
the 2HDM II. Stringent constraints are thus obtained on the
parameter space of the model. The results are summarized as
follows.  In the 2HDM II, the dominant contribution actually comes
at the two-loop level, and in order to explain the data $\Delta
a_\mu$ is preferred to come from the Higgs pseudoscalar  $A$, the
mass of which is required to be less than 29 (55) (85) GeV for
$\tan\!\beta=30\, (45)\, (60)$ by the 95\% C.L. range of
Eq.(\ref{95}). Moreover, $\tan\!\beta$ has to be larger than 17.
Recently, there has been some work on the same subject
\cite{2hdm}; however, the paper only considered the one-loop
contributions from the Higgs bosons, which become smaller than their
corresponding two-loop contributions when the Higgs bosons are heavier
than a few GeV and substantially smaller for heavier Higgs bosons. It
is important to note that the two-loop contributions considered
here dramatically change the story of the Higgs sector
contributions to $a_\mu$ and invalidate most of the results of the
one-loop studies. Although we focus on the 2HDM II here, a similar
conclusion holds for most models with possible large contributions
from the Higgs sector. In models with flavor-changing Higgs
couplings, in particular, there are potentially 1-loop
contributions substantially larger than the flavor-conserving
ones\cite{ohm}. However, the Barr-Zee type 2-loop contributions
would still have an important role to play and should be taken
into consideration. This fact has often been overlooked in the
literature.

In the MSSM, the dominant contribution comes from the chargino-sneutrino-loop
diagrams and it has been shown \cite{susy} that to satisfy $\Delta a_\mu$
requires the gaugino mass and the smuon mass below about $600-800\,\mbox{GeV}$.
Because of additional mass constraints on the scalars in MSSM,
the total contribution from the Higgs bosons is not at a significant level and thus
will not affect the conclusion of Ref.\cite{susy}. While the typical failure
of the MSSM studies to address this kind of 2-loop contribution is a potential
problem, our results here get rid of the worry, at least for the case of a more
generic scalar mass spectrum.

Many other extensions of the SM have extra contributions to the $a_\mu$. Some
examples are additional gauge bosons\cite{Z},  leptoquarks\cite{lq}, and
muon substructure\cite{other}. However, not all of them can contribute
in the right direction as indicated by the data.  Thus, the
$a_\mu^{\rm exp}$ measurement can differentiate among various models,
and perhaps with other existing data can put very strong constraints
on the model under consideration.

\begin{figure}
\includegraphics{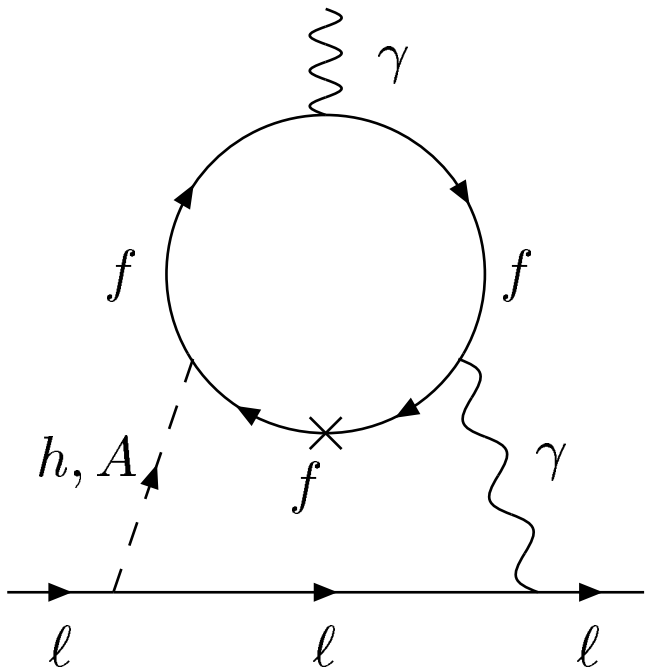}
\vspace*{6in}
\caption{ \label{fig1}
The dominant two-loop graph involving a scalar or
a pseudoscalar boson that contributes to $a_\ell$.  }
\end{figure}

Given the mass bound on SM Higgs boson \cite{lep}, the Higgs contribution
to $\Delta a_\mu$ at one-loop level is negligible.  However, it has been
emphasized, in Ref.\cite{2loop} for example, that for Higgs boson mass larger than 
about $3\,\mbox{GeV}$, the dominant Higgs contribution to  $a_\mu$ actually comes
from the two-loop Barr-Zee diagram (first discussed by Bjorken and Weinberg)
\cite{BZ} with a heavy fermion ($f$)
running in the loop. A $m_f^2/m_\mu^2$ factor could easily overcome the
$\alpha/4\pi$ loop factor. The two-loop scalar contribution with a
heavy fermion $f$, as shown in Fig.~\ref{fig1}, is given by
\begin{equation} \label{s}
\Delta a_\ell^{h} = -
\frac{N_{\!c}^f \, \alpha^2 }{4\pi^2 \sin\!^2\theta_{\! \scriptscriptstyle W}}
\frac{m_\ell^2 \, \lambda_\ell}{M_{\! \scriptscriptstyle W}^2}\;
{\cal Q}_f^2 \, \lambda_f \,
f\!\!\left( \frac{m_f^2}{m_h^2} \right) \; ,
\end{equation}
where
\beq
f(z)={1\over 2} z \int^1_0 \! dx \; \frac{1-2x(1-x)}{x(1-x)-z} \ln\frac{x(1-x)}{z} \; ,
\eeq
$N_{\!c}^f$ represents the number of color degrees of freedom in $f$, and
${\cal Q}_f$ its electric charge.  Here $\ell$ denotes a generic
lepton, $m_h$ is the (scalar) Higgs boson mass, and $\lambda_\ell$ and
$\lambda_f$ represent plausible modifications to the Higgs couplings of the
fermions ($\lambda_\ell = \lambda_f =1$ in the SM).

Ref.\cite{CM} quoted an electroweak contribution, calculated up to two-loop
level, of
\begin{equation}
a_\mu^{\rm EW} = 152 (4) \; \times \; 10^{-11} \;.
\end{equation}
Included in this number is a Barr-Zee diagram contribution with a
$t$ loop. The numerical value of this result barely exceeds the order
of $10^{-11}$ (see also Ref.\cite{CKM1}) and is negative for any
reasonable value of $m_h$. Moreover, there are other purely bosonic
two-loop contributions \cite{CKM2}, in which the SM Higgs boson also plays a
role.  Nevertheless, all the contributions involving the SM Higgs boson are
quite small\cite{CM,CKM1,CKM2}.
When considering a model with an extended Higgs sector, a complete analysis
would first require one to subtract the SM Higgs contribution and recalculate
all the Higgs contributions. This is because the number of Higgs bosons,
their effective couplings, and mass constraints would be different from the SM
scenario. Nevertheless, in our study here, we only calculate the Higgs
Barr-Zee diagram contributions and assume that this does give a very
good approximation of $a_\mu$ from the diagrams involving Higgs bosons
($\Delta a_\mu^{\rm \tiny Higgs}$). Our rationale is as follows.  We
are interested in the region of parameter space where the Higgs
contributions could explain the discrepancy of Eq.(\ref{exp}), or at
least, in the case of models that have other important contributions, do
play a substantial role in $\Delta a_\mu$. Hence, we focus on the
region where $\Delta a_\mu^{\rm \small Higgs}$ is at or close to the
order of $10^{-9}$.  As we will see below, the possibility of having
such largely enhanced Higgs contributions comes from the combined
effect of coupling enhancements and weakened Higgs mass constraints.
The coupling enhancement is only to be found among the
Yukawa couplings, thus illustrating the special importance of the Barr-Zee
diagram considered.

In a model with an extended Higgs sector, we can write the fermion
couplings of a neutral Higgs mass eigenstate $\phi^0$ as
\begin{equation}
{\cal L}^{\footnotesize \bar{f}\!\phi^0\! f}
=  - \lambda_{f} \, \frac{g \,m_{\! \scriptscriptstyle f}}{2 M_{\! \scriptscriptstyle W}}
\bar{f}\phi^0 f +
i \gamma_{\scriptscriptstyle 5}\, A_{f}  \,
\frac{g \,m_{\! \scriptscriptstyle f}}{2 M_{\! \scriptscriptstyle W}}
\bar{f} \phi^0 f \; ,
\end{equation}
where $\lambda_{f}\,
\frac{g \,m_{\! \scriptscriptstyle f}}{2 M_{\! \scriptscriptstyle W}}$ and
$A_{f}\, \frac{g \,m_{\! \scriptscriptstyle f}}{2 M_{\! \scriptscriptstyle W}}$ are
the effective scalar and pseudoscalar couplings.  The contribution of two-loop
diagram (Fig.~\ref{fig1}) to $a_\mu$ is then given by the sum of Eq.(\ref{s})
(with $m_h=m_{\phi^0}$) and the pseudoscalar expression
\begin{equation} \label{ps}
\Delta a_\ell^{\! \scriptscriptstyle A} =
\frac{N_{\!c}^f \, \alpha^2 }{4\pi^2 \sin\!^2\theta_{\! \scriptscriptstyle W}}
\frac{m_\ell^2 \, A_\ell}{M_{\! \scriptscriptstyle W}^2}\;
{\cal Q}_f^2 \, A_f \,
g\!\!\left( \frac{m_f^2}{m_{\! \scriptscriptstyle A}^2} \right) \;
\end{equation}
(with $m_{\! \scriptscriptstyle A}=m_{\phi^0}$), where
\beq
g(z)={1\over 2} z \int^1_0 \!  dx \; \frac{1}{x(1-x)-z} \ln\frac{x(1-x)}{z} \;.
\eeq
Without CP violation in the Higgs sector, Eqs.(\ref{s}) or (\ref{ps}) gives
directly the contribution from a scalar or a pseudoscalar, respectively.
The corresponding contributions with the $\gamma$ in Fig.~\ref{fig1}
replaced by the $Z^0$ are suppressed by about two orders of
magnitude\cite{2loop}, hence neglected here. We also skip the details
about the similar contribution from a charged Higgs boson, which involves a
$W^\pm$ boson and is thus suppressed also. Note that a light charged Higgs 
boson less than $80.5\,\mbox{GeV}$ is ruled out by the 
CERN $e^+ e^-$ collider (LEP) experiments\cite{lep}.
Moreover, analysis of its contribution to $b\to s\,\gamma$ leads to a much
stronger lower bound --- $380\,\mbox{GeV}$, as claimed in Ref.\cite{bsg},
for instance.

The 2HDM has three physical neutral Higgs bosons: two scalars $h$ and $H$, and
the pseudoscalar $A$. For model II under consideration, the corresponding nonzero
$\lambda_{f}$ or $A_{f}$ for $f= t, b$, and $\ell (=e, \mu, \tau)$ are given by
\beqa
\qquad && h (\lambda_{f}) : \qquad \frac{\cos\!\za}{\sin\!\zb}
    \qquad -\frac{\sin\!\za}{\cos\!\zb} \qquad -\frac{\sin\!\za}{\cos\!\zb}
\nonumber \\
\qquad  && H (\lambda_{f}) : \qquad \frac{\sin\!\za}{\sin\!\zb}
    \qquad \;\;\frac{\cos\!\za}{\cos\!\zb} \qquad \;\;\;\;\frac{\cos\!\za}{\cos\!\zb}
\nonumber \\
\qquad  && A (A_{f}) : \qquad {\cot\!\zb} \qquad \;\;\;{\tan\!\zb} \qquad \;\;\;\;\,{\tan\!\zb}
\nonumber
\eeqa
respectively, in the standard notation\cite{guide}. What is particularly
interesting phenomenologically is the enhancement of the couplings of
$\bar{b} b H,  \bar{b} b h, \bar\ell \ell H,  \bar\ell \ell h, \bar b b A,
\,\mbox{and} \, \bar\ell \ell A$ at large $\tan\!\zb$.  In fact, the
dominant contributions then come from the diagrams with a $b$ or $\tau$ loop.

It was pointed out in Ref.\cite{2loop} that the two-loop
pseudoscalar contribution to $a_\mu$ is positive while the two-loop
scalar contribution is negative
in the large $\tan\!\beta$ region. Note that this is always true for the
dominating contributions with a $b$ or $\tau$ loop, independent of
the scalar mixing angle $\alpha$. The reverse happens in the
corresponding one-loop contributions, but these one-loop contributions are
suppressed relative to the two-loop contributions.
In fact, using the one-loop $a_\mu$ result in constraining the 2HDM II has
been studied extensively\cite{1loop}.  Using the two-loop result, however,
changes the story dramatically and invalidates most of the conclusions from
the one-loop studies.

\begin{figure}[th]
\centering
\includegraphics[width=5.2in]{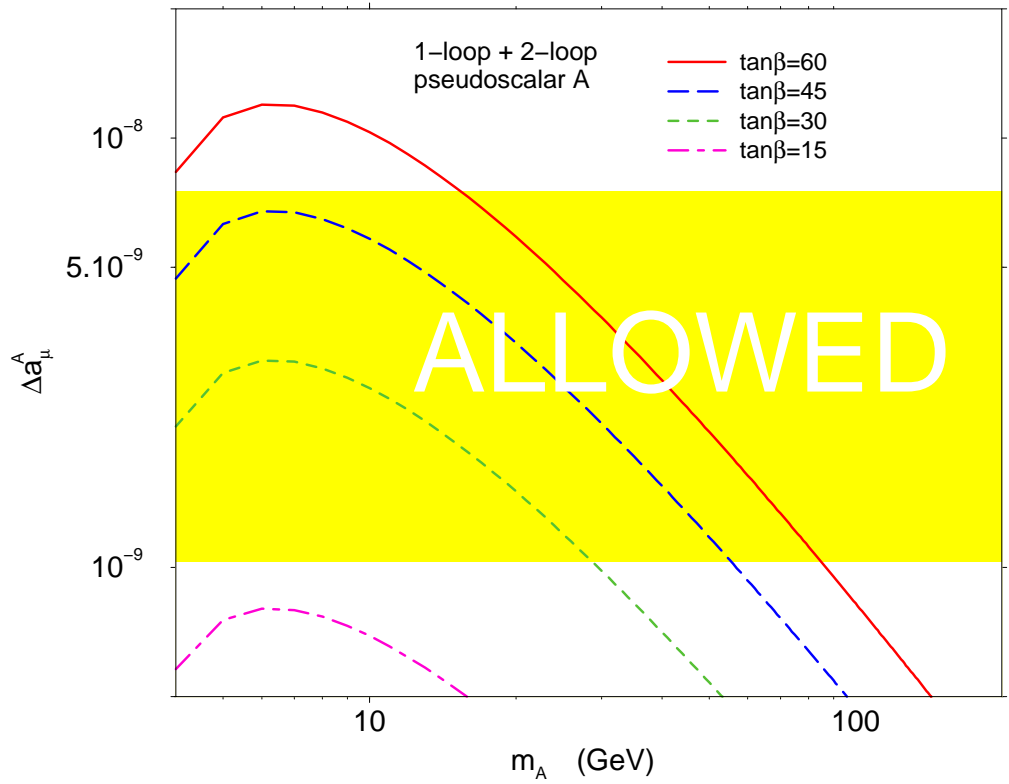}
\medskip
\caption{ \label{fig2}
$\Delta a_\mu^A$ (one-loop and two-loop summed)
versus the pseudoscalar Higgs boson mass
$m_{\!\scriptscriptstyle A}$ for various values of $\tan\!\beta$.}
\end{figure}

Applying the $\Delta a_\mu$ constraint to the 2HDM II, we need to suppress the
scalar-Higgs ($h$ and $H$) contributions, as it comes 
in the opposite direction as indicated by the data, relative to the
pseudoscalar contribution.  This is in direct contradiction to what is suggested
in the one-loop studies. At large $\tan\!\beta$ both the dominating contributions
from the scalars and the pseudoscalar scale roughly as $\tan\!\beta$.
For the scalar part, we can adjust the mixing angle $\alpha$ to zero
such that the contribution from the light Higgs boson $h$ is negligible, and
impose a large mass hierarchy between the scalar Higgs $H$ and the
pseudoscalar $A$.  Then a relatively light mass for the pseudoscalar $A$ will
give a sufficiently large positive contribution to $\Delta a_\mu$, or the
required $\Delta a_\mu$ value could be used to obtain the admissible range for
$m_{\!\scriptscriptstyle A}$. In Fig.~\ref{fig2}, we show the contribution of
$\Delta a_\mu^A$ from the pseudoscalar $A$ versus $m_{\!\scriptscriptstyle A}$
for various values of $\tan\!\beta$.  We included the one-loop and two-loop
pseudoscalar  contributions. The shaded region is the 95\% C.L. range of
Eq.(\ref{95}). The required range of $m_{\!\scriptscriptstyle A}$
is then given by about 4 -- 29 (15 -- 85) GeV for $\tan\!\beta= 30\, (60)$.
Moreover, a $\tan\!\beta \geq 17$ is always required.

\begin{figure}[th]
\centering
\medskip
\includegraphics[width=3.2in]{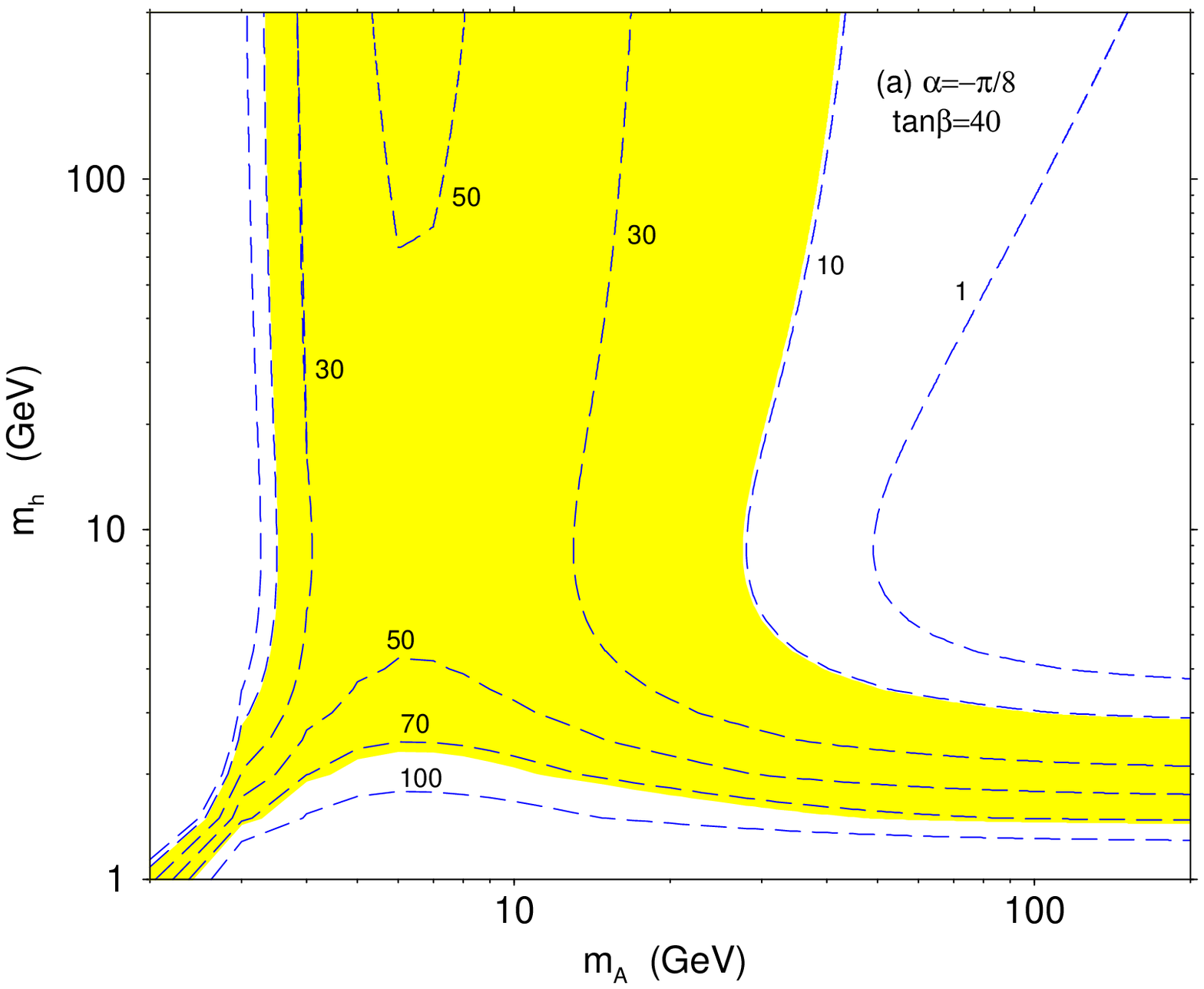}
\includegraphics[width=3.2in]{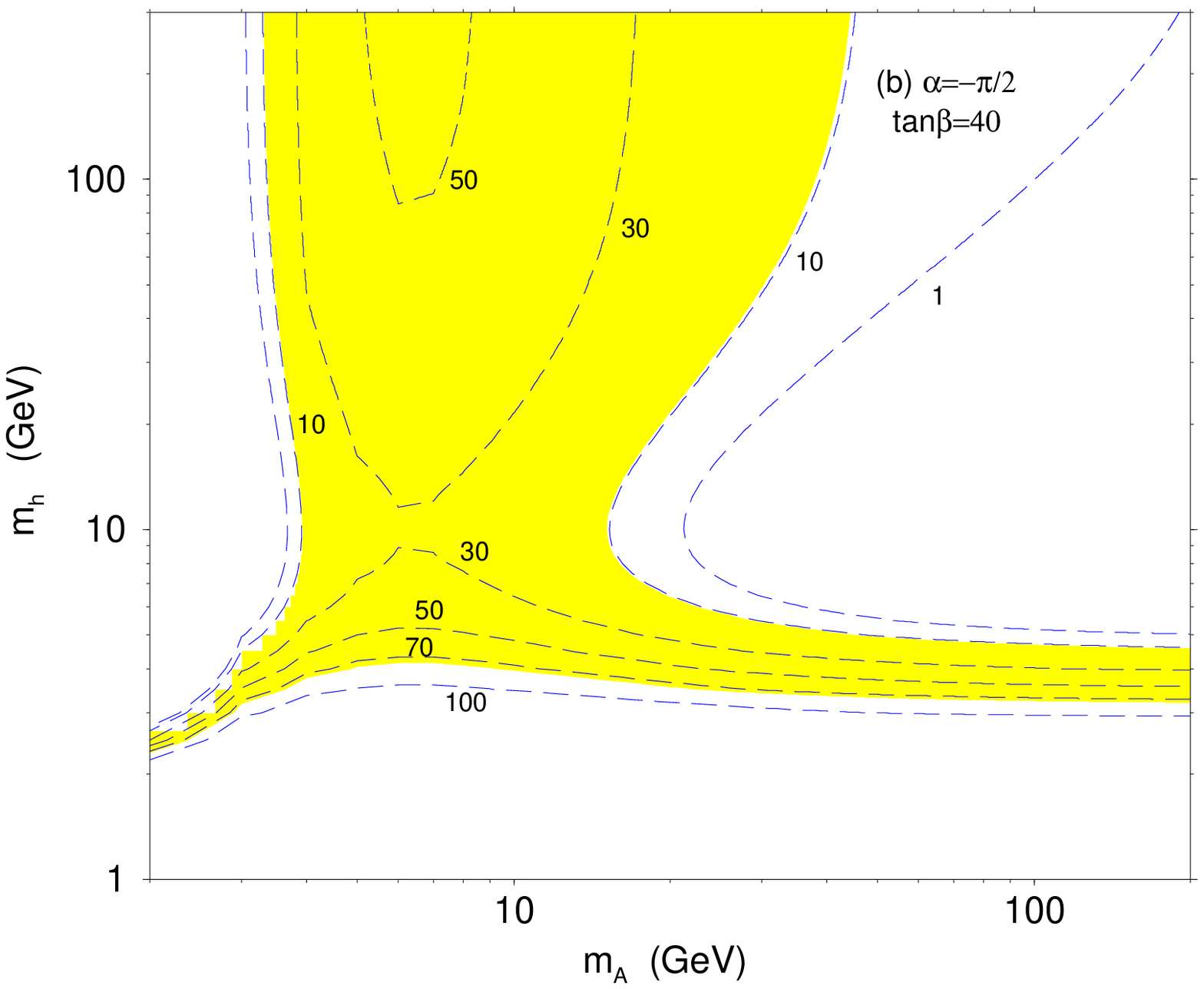}

\includegraphics[width=3.2in]{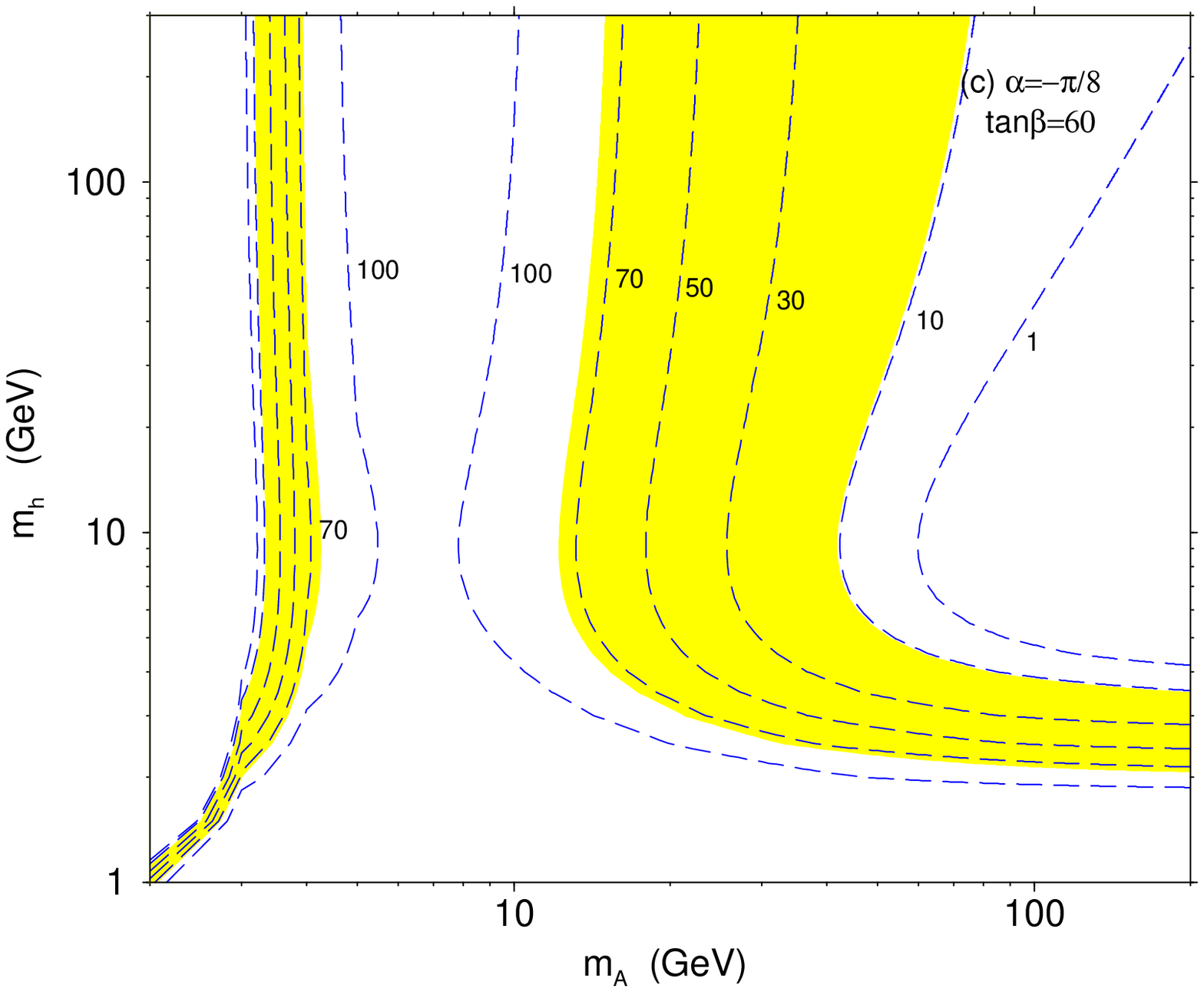}
\includegraphics[width=3.2in]{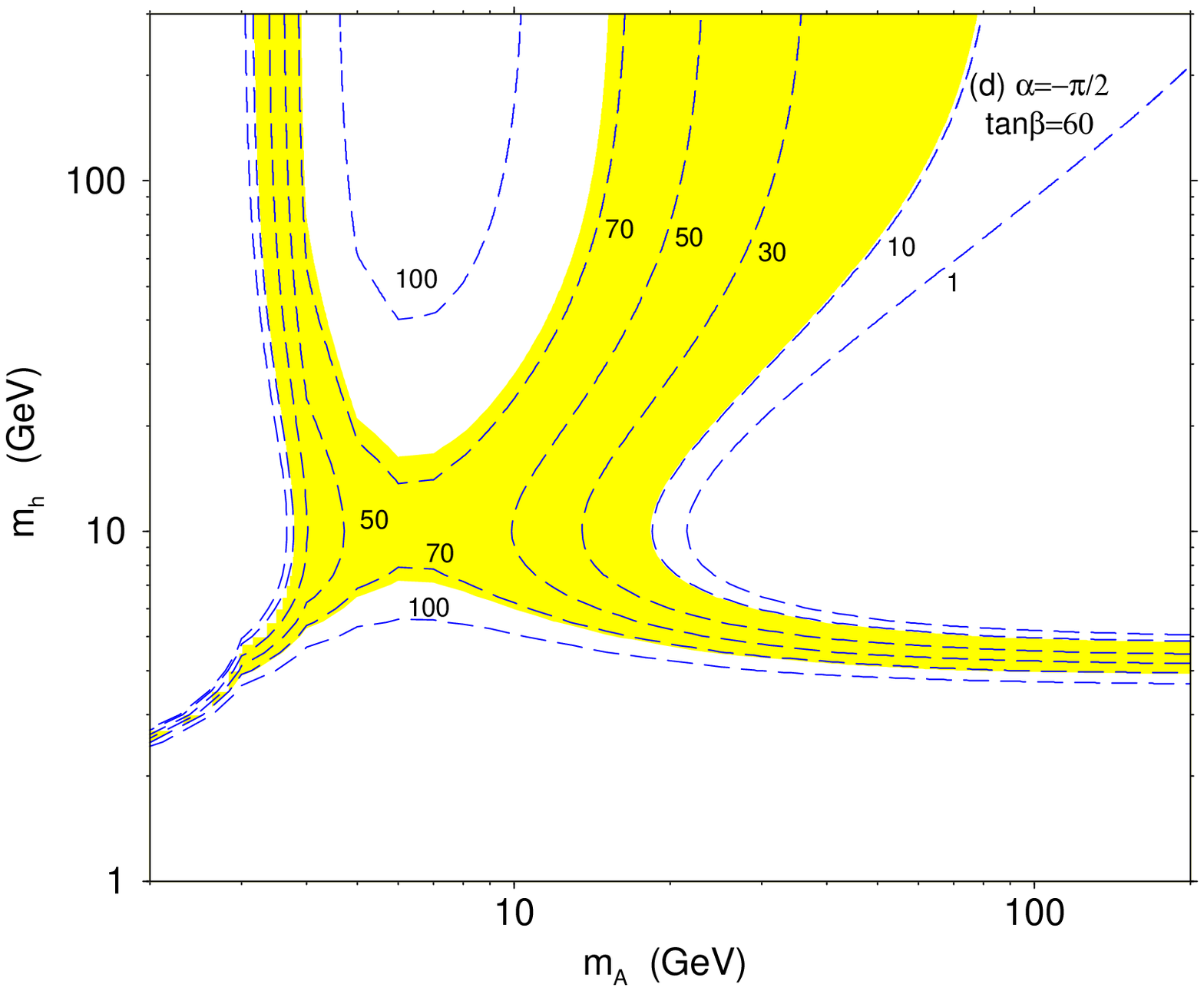}
\medskip
\caption{ \label{fig3}
Contours of $\Delta a_\mu^{\rm\tiny Higgs}$ in unit of $10^{-10}$ in the
plane of $(m_{\!\scriptscriptstyle A},m_h)$ for
(a) $\alpha=-\pi/8$ and  $\tan\!\beta=40$,
(b) $\alpha=-\pi/2$ and  $\tan\!\beta=40$,
(c) $\alpha=-\pi/8$ and  $\tan\!\beta=60$, and
(d) $\alpha=-\pi/2$ and  $\tan\!\beta=60$.
}
\end{figure}

What happens when other mixing angles $\alpha$ are chosen?  The light Higgs
boson $h$ will give a negative contribution to $a_\mu$ and thus offsets the
pseudoscalar contribution.  Therefore, the required range of
$m_{\!\scriptscriptstyle A}$ shifts to a lower value in order to accommodate
the data.  We show the contour plots of $\Delta a_\mu$ in the plane of
$(m_{\!\scriptscriptstyle A},m_h)$ for a small and a large  $\alpha$ at
$\tan\!\beta= 40$ and $60$ in Fig.~\ref{fig3}. The $\alpha=0$ limit, which
corresponds to switching off the contribution from the scalar $h$, can be
easily read off  from the vertical asymptotes. On the other hand,
$\alpha= \pm \pi/2$ corresponds to the maximal contribution from $h$. The
contour plots show how the contribution from the scalar $h$ affects the solution
to $a_\mu$. For example, at $\alpha=-\pi/2$ for $m_h \sim 60\,\mbox{GeV}$,
the 95\% C.L. required range of $m_{\!\scriptscriptstyle A}$ lowers to
13--47 GeV for $\tan\!\beta=60$.

At this point, it is interesting to take into consideration other experimental
constraints on Higgs masses.  A collider search for the neutral Higgs bosons
in the context of the 2HDM typically rules out a region of small $m_h$ and
$m_{\! \scriptscriptstyle A}$. In particular, an OPAL analysis\cite{OPAL} using
the LEP II data up to $\sqrt{s}=189\,\mbox{GeV}$ excludes the regions
$1 < m_h < 44\,\mbox{GeV}$ {\it and}
$12 <  m_{\! \scriptscriptstyle A}< 56\,\mbox{GeV}$ at 95\% C.L.,
independent of $\alpha$ and $\tan\!\beta$. This is a conservative limit.
Details of the exclusion region vary with $\alpha$ and $\tan\!\beta$, and go
substantially beyond the rectangular box\cite{OPAL}. Essentially, the search for
$A$ relies on the process $e^+ e^- \to A \, h$, therefore, if $m_h$ is so
large that this process becomes negligible for $A$ production, there would be no
limit on $m_{\!\scriptscriptstyle A}$.  In addition, below
$m_{\!\scriptscriptstyle A} \simeq 5$ GeV, the direct search in $e^+ e^- \to A\,h$
was not included because the detection efficiency vanishes and the total $Z^0$
width only provides very limited exclusion. The OPAL exclusion region is
roughly sitting at the center on the $m_{\!\scriptscriptstyle A}$ axis going
up to about $m_h = 60\,\mbox{GeV}$ in the plots of Fig.~\ref{fig3}, and cuts
out part of the admissible region of the $\Delta  a_\mu$ solution. For $m_h$
larger than the OPAL limit the admissible $m_{\!\scriptscriptstyle A}$ range is
roughly from 3 to 50 GeV for $\tan\!\beta=40$. At larger $\tan\!\beta$
(60 as the  illustrated example), the range widens, especially at the upper end;
but a middle range (about 4 to 15 here) is lost as $\Delta  a_\mu$ gets too
large. There is another admissible small window at very small $m_h$ (a few GeV)
with large $m_{\! \scriptscriptstyle A}$. This region is indeed dominated
by the one-loop contribution from $h$, but even here, the two-loop contribution
(from $h$) has an important role to play. This tiny $m_h$ window
and the similar solution with this kind of small $m_{\! \scriptscriptstyle A}$ are
in fact excluded by Upsilon decay\cite{U} and some other processes\cite{guide}.

In addition to the constraint from direct search, there are also
other constraints on the masses of the Higgs bosons coming from
the electroweak precision data. While a comprehensive treatment of
the topic is really beyond the scope of the present study, we
discuss below the basic features, using results from a recent
paper\cite{take}. The scope of the latter study is limited to the
large $\tan\beta$ region and  $\alpha=\beta \approx \pi/2$. There,
the ratio $m_h/m_{\! \scriptscriptstyle A}$ is constrained at 95\%
C.L. (based on the Bayesian approach) via the function
\begin{displaymath}
G \left ( \frac{m_h^2}{m_{\! \scriptscriptstyle A}^2} \right ) \ge
- \left( \frac{39}{\tan\!\beta}
 \right )^2 \;,
\end{displaymath}
where
\begin{displaymath}
G(x) = 1 + \frac{1}{2} \left( \frac{1+x}{1-x} \right )\, \ln x \;.
\end{displaymath}
Solving for $m_h/m_{\! \scriptscriptstyle A}$ at $\tan\!\beta =
60,45, 30$, we obtain, respectively,
\begin{eqnarray}
0.3 \qquad  \le & \qquad \frac{m_h}{m_{\! \scriptscriptstyle A}}
\qquad & \le \qquad 3.2\nonumber \\ 0.2  \qquad \le & \qquad
\frac{m_h}{m_{\! \scriptscriptstyle A}} \qquad & \le \qquad 5.1
\nonumber \\ 0.07 \qquad \le &\qquad \frac{m_h}{m_{\!
\scriptscriptstyle A}} \qquad & \le \qquad 14.4\nonumber \;.
\end{eqnarray}
We can see that the precision electroweak data prefers a region
close to the diagonal of the $m_h$ verses $m_{\!
\scriptscriptstyle A}$ plot. If one naively imposes the result of
$0.3 \le \frac{m_h}{m_{\! \scriptscriptstyle A}}  \le 3.2$ at
$\tan|!\beta=60$ onto  Fig. 3(d), which is at a different
$\alpha$ but with which a similar result is expected to be valid,
together with the direct search limit $m_h \agt 60$ GeV and the
$a_\mu$ requirement, only a small ``triangle" is left. This
triangle is bounded by $m_h=60$ GeV, $m_h/m_{\! \scriptscriptstyle
A}=3.2$ and the contour of $a_\mu=10\times 10^{-10}$ [labeled by
``10" in Fig. 3(d)]. The surviving parameter space region,
however, is in the more favorable ``larger" mass area. In
particular, it re-enforces our previous comment at the end of the
last paragraph that the one-loop dominating tiny $m_h$ window of
solution to $a_\mu$ is ruled out. The situation for the other $\alpha$
values is expected to be similar. It will be very interesting to
have the complete phenomenological analysis combining all the
constraints on the 2HDM.

Finally, we discuss the role of the Higgs sector contributions to $a_\mu$
in the MSSM. The LEP bound on the Higgs boson masses is in the range
85 -- 95 GeV\cite{spm}. From the above result, one may naively conclude that
if a Higgs particle is just around the corner, it could have an important role
to play in $a_\mu$. However, there are some strong theoretical constraints on
the relation of the Higgs boson masses in the MSSM. At the large $\tan\!\beta$ value
required, one has $m_h \lsim m_{\!\scriptscriptstyle A}
\simeq m_{\!\scriptscriptstyle H}$ \cite{guide}. Most of the Higgs contributions 
to $a_\mu$ cancel among themselves. Moreover, a small Higgs boson mass
may require a $\mu$ parameter so small that the chargino/neutralino contributions
to $a_\mu$ get far too large. In fact, we have checked and found
no interesting solution within MSSM in which the Higgs contributions play a
substantial role. In the admissible range of chargino/smuon masses found in
Ref.\cite{susy},  the total Higgs contribution is only about 1\% of the SUSY 
contribution. The importance of this null result should not be underestimated.

We conclude that the new measurement on the muon anomalous magnetic moment
constrains severely on the parameter space of the 2HDM II, and our results,
including one-loop and two-loop contributions, change dramatically from the
conclusion that one draws by using only one-loop results.

\section*{Acknowledgement}
This research was supported in part by the National Center for
Theoretical Science under a grant from the National Science
Council of Taiwan R.O.C. We thank A. Arhrib, Stephen Narison, John
Ng, and T. Takeuchi for useful discussions. One of us (C.-H. Chou)
is particularly in debt to D. Chang, W.-F. Chang, and W.-Y. Keung
for the previous collaboration on the topic.



\vspace*{2in}

\noindent{\bf Figure Captions :-}
\\[.3in]
Fig.~1 --- The dominant two-loop graph involving a scalar or
a pseudoscalar boson that contributes to $a_\ell$.
\\[.2in]
Fig.~2 --- $\Delta a_\mu^A$ (one-loop and two-loop summed)
versus the pseudoscalar Higgs boson mass
$m_{\!\scriptscriptstyle A}$ for various values of $\tan\!\beta$.
\\[.2in]
Fig.~3 --- Contours of $\Delta a_\mu^{\rm\tiny Higgs}$ in unit of $10^{-10}$
in the
plane of $(m_{\!\scriptscriptstyle A},m_h)$ for
(a) $\alpha=-\pi/8$ and  $\tan\!\beta=40$,
(b) $\alpha=-\pi/2$ and  $\tan\!\beta=40$,
(c) $\alpha=-\pi/8$ and  $\tan\!\beta=60$, and
(d) $\alpha=-\pi/2$ and  $\tan\!\beta=60$.

\newpage

\eject
\vspace*{1in}

\eject
\vspace*{1in}

\end{document}